%% file: nierste_discrete24.tex
\documentclass[a4paper,11pt]{article}
\usepackage{pos}
\usepackage{color}
\input{def_colours}

\usepackage{amsmath}
\usepackage{cancel}
\newcommand{\eq}[1]{Eq.\,(\ref{#1})}
\newcommand{\eqsand}[2]{Eqs.\,(\ref{#1}) and (\ref{#2})}
\newcommand{\eqsto}[2]{Eqs.\,(\ref{#1}) to (\ref{#2})}

\renewcommand{\thefootnote}{\fnsymbol{footnote}}
\def\beq#1\eeq{\begin{align}#1\end{align}}

\newcommand{\fig}[1]{Fig.~\ref{#1}}
\newcommand{\ov}{\overline}

\newcommand{\eg}{{\em e.g.}}
\newcommand{\ie}{{\em i.e.}}

\newcommand{\ket}[1]{\left\vert{#1}\right\rangle}
\newcommand{\no}{\nonumber}
\newcommand{\nn}{\nonumber\\}

\newcommand{\DorDbar}{\raisebox{6.9pt}{$\scriptscriptstyle\!(\hspace*{6.9pt})$}
  \hspace*{-11.4pt}\bar D \,{}^0\,}

\usepackage{hyperref}
\newcommand{\arxhref}[1]{\href{https://arxiv.org/abs/#1}{#1}}  

\title{Branching fractions and CP asymmetries in charm meson decays}

\author*{Ulrich Nierste}

\affiliation{Institute for Theoretical Particle Physics (TTP), Karlsruhe
  Institute of Technology (KIT)\\
  Wolfgang-Gaede-Stra\ss e 1, 76131 Karlsruhe, Germany}

\emailAdd{ulrich.nierste@kit.edu}

\abstract{I present a consistent way to include $\eta$-$\eta^\prime$
  mixing in global analyses of two-body decays of heavy hadrons employing
 the approximate flavour-SU(3) symmetry of QCD. The framework is applied
 to $D\to P \eta^\prime$ decays, where $P$ denotes a pseudoscalar meson.
 The result shows that flavour-SU(3) symmetry holds in the decay rates
 of these modes to better than 30\%. With future data we expect the
 branching ratios of $D_s\to K^+ \eta^\prime$ and $D \to K^+
 \eta^\prime$  to move upward and downward by $\sim\!\! 1\sigma$, respectively. 
 Subsequently I discuss the implications of the LHCb measurements of
 the CP asymmetries in $D\to K^+K^-$ and $D\to \pi^+\pi^-$ for generic
 scenarios of new physics. New-physics contributions should have
 imprints on other CP asymmetries as well and can be tested through
 sum rules. Promising decays are $D_s^+\to K^0\pi^+$,
 $D^+\to \bar K^0K^+$, $D^0\to K^0 \bar K^{*0}$,
 $D^0\to \bar K^0 K^{*0}$, $D_s^+\to K^{*0}\pi^+$, and
 $D^+\to \bar K^{*0}K^+$.}

\FullConference{9th Symposium on Prospects in the Physics of Discrete Symmetries (DISCRETE2024)\\
 2–6 Dec 2024\\
Ljubljana, Slovenia\\}

\begin{document}
\maketitle

\section{Overview}
In the Standard Model (SM) flavour-changing transitions are encoded in
the Cabibbo-Kobayashi-Maskawa (CKM) matrix
\begin{eqnarray}
\begin{pmatrix}
\displaystyle  \color{BrickRed}V_{ud} &\displaystyle   \color{BrickRed}V_{us} 
&\displaystyle \color{Blue}V_{ub} \\
\displaystyle \color{BrickRed} V_{cd} 
&\displaystyle  \color{BrickRed}V_{cs} &\displaystyle  \color{Blue}V_{cb} \\
\displaystyle V_{td} &\displaystyle  V_{ts} 
&\displaystyle  V_{tb}
\end{pmatrix} 
& \simeq &
\begin{pmatrix}
1-\frac{{\color{BrickRed}\lambda^2}} {2} &{  \color{BrickRed}\lambda} &
   \!\!  {\color{Blue} A} { {\color{BrickRed}\lambda^3}}
 ({\color{OliveGreen}\rho}  - i {\color{OliveGreen} \eta} ) \\
- \color{BrickRed}\lambda 
& 1 -  \frac{ {\color{BrickRed}\lambda^2} }{2} &
{ \color{Blue}A } \color{BrickRed}\lambda^2  \\
{\color{Blue}A }{\color{BrickRed}\lambda^3}
(1 - {\color{OliveGreen} \rho} - i 
 {\color{OliveGreen}\eta} )
& - \color{Blue}A \color{BrickRed}\lambda^2  & 1
\end{pmatrix}
\end{eqnarray}
with Wolfenstein parameters
${\color{BrickRed}\lambda},{\color{Blue}A},{\color{OliveGreen}\rho},{\color{OliveGreen}\eta}$.
Charm decays involve the red and blue CKM elements and
have no stakes in Standard-Model (SM) CKM metrology. Yet they play a unique
role in probing new physics in the flavour sector of up-type quarks.

\begin{figure}
\begin{center}\includegraphics[height=2cm]{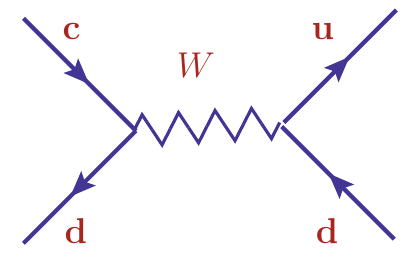}
\hspace{10mm}\includegraphics[height=2cm]{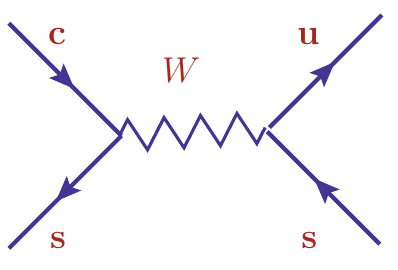}
\hspace{15mm}
\includegraphics[height=2cm]{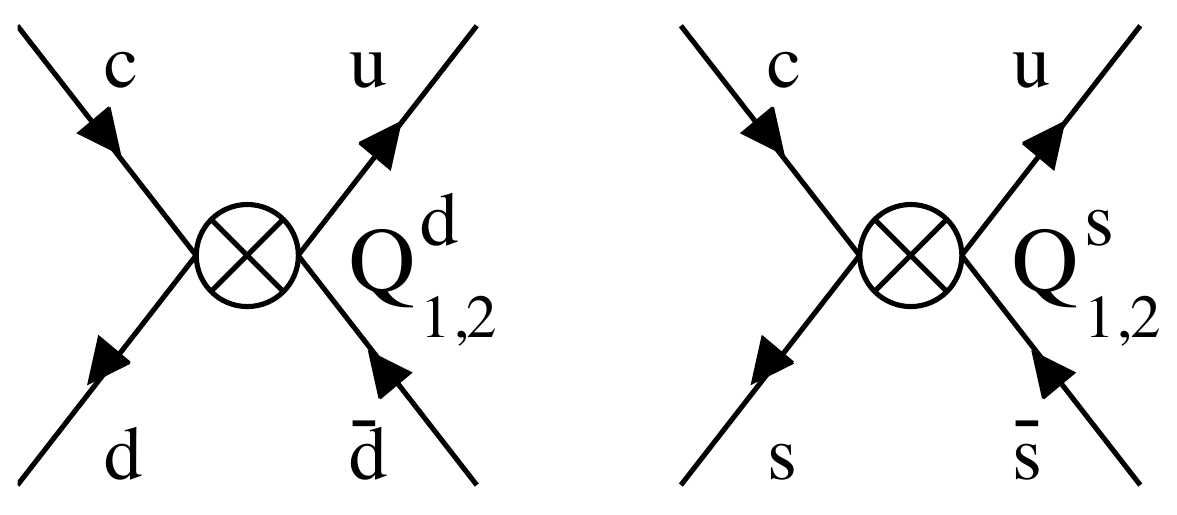}
\end{center}
\caption{Tree-level contribution to singly Cabibbo-suppressed
  (SCS) charm decays in the Standard Model and the weak effective theory. 
  \label{fig:tree}}
~\\[-3mm]
\hrule
\end{figure}

I discuss decays of $D^0,D^+,D_s^+$ mesons into two pseudoscalar mesons,
$D\to P P^\prime$, or a pseudoscalar and a vector meson, $D\to PV$. All
these decays are dominated by a $W$-mediated tree amplitude,
categorised by the power of the Wolfenstein parameter $\lambda=0.225$:
\begin{itemize}
\addtolength{\itemsep}{-5pt}
\item Cabibbo-favoured (CF), ${\cal O}(\lambda^0)$, $c\to s \bar d
  u$.
\item Singly Cabibbo-suppressed (SCS),   ${\cal O}(\lambda^1)$, $c\to d \bar d
  u$ or $c\to s \bar s  u$.
\item Doubly   Cabibbo-suppressed (DCS),   ${\cal O}(\lambda^2)$, $c\to d \bar
  s u$. 
\end{itemize}
The tree diagrams for SCS decays are shown in \fig{fig:tree}. Since the
energy scale of $D$ meson decays is way below the mass of the $W$ boson,
we can describe these decays by a point-like four-fermion interaction in
analogy to the Fermi interaction, resulting in the operators $Q_2^d$ and
$Q_2^s$ with
$Q_2^q\equiv \bar q_L^\alpha \gamma_\mu c_L^\alpha \, \bar u_L^\beta
\gamma^\mu q_L^\beta$. Here $\alpha$ and $\beta$ are colour indices.
The resulting weak effective theory is set up to accomodate Quantum
Chromodynamics (QCD) effects, which requires to include the
colour-flipped operators
$Q_1^q\equiv \bar q_L^\alpha \gamma_\mu c_L^\beta \, \bar u_L^\beta
\gamma^\mu q_L^\alpha$ in our weak effective lagrangian.

Branching fractions in $D\to PP ^\prime$  or $D\to PV$ decays are insensitive to
new physics and are ``bread and butter'' physics to test the calculational tools
and check the data for consistency. On the contrary, CP asymmetries are
tiny in the SM and thus very sensitive to new physics. SCS decays
involve
\begin{eqnarray}
\lambda_d&=&V_{cd}^*V_{ud},\qquad
\lambda_s\; =\; V_{cs}^*V_{us},\qquad
\lambda_b\; =\; V_{cb}^*V_{ub}. 
\end{eqnarray}
Note that $|\lambda_d|\simeq |\lambda_s|\gg  |\lambda_b|$. By using
CKM unitarity, $\lambda_d=-\lambda_s-\lambda_b $, one verifies that
\emph{all}\ SM CP asymmetries are proportional to
\begin{eqnarray}
  \mbox{Im}\,\frac{\lambda_b}{\lambda_{s}}&=& -6\cdot 10^{-4}.
                                              \label{imbs}
\end{eqnarray}
Therefore even in decays in which the two large tree-level amplitudes
$c\to d \bar d u$ and $c\to s \bar s u$ interfere, the resulting CP
asymmetry involves the suppression factor in \eq{imbs} owing to
$ \mbox{Im}\,\frac{\lambda_d}{\lambda_{s}}=-
\mbox{Im}\,\frac{\lambda_b+\lambda_s}{\lambda_{s}} = -
\mbox{Im}\,\frac{\lambda_b}{\lambda_{s}}$. Prominent sample decays for
this tree-tree interference are
$D^0\to K_S K_S$ and $D^0\to K_S K^{*(0)}$ 
\cite{Nierste:2015zra,Nierste:2017cua}.

\begin{figure}
\centering\includegraphics[width=0.6\textwidth]{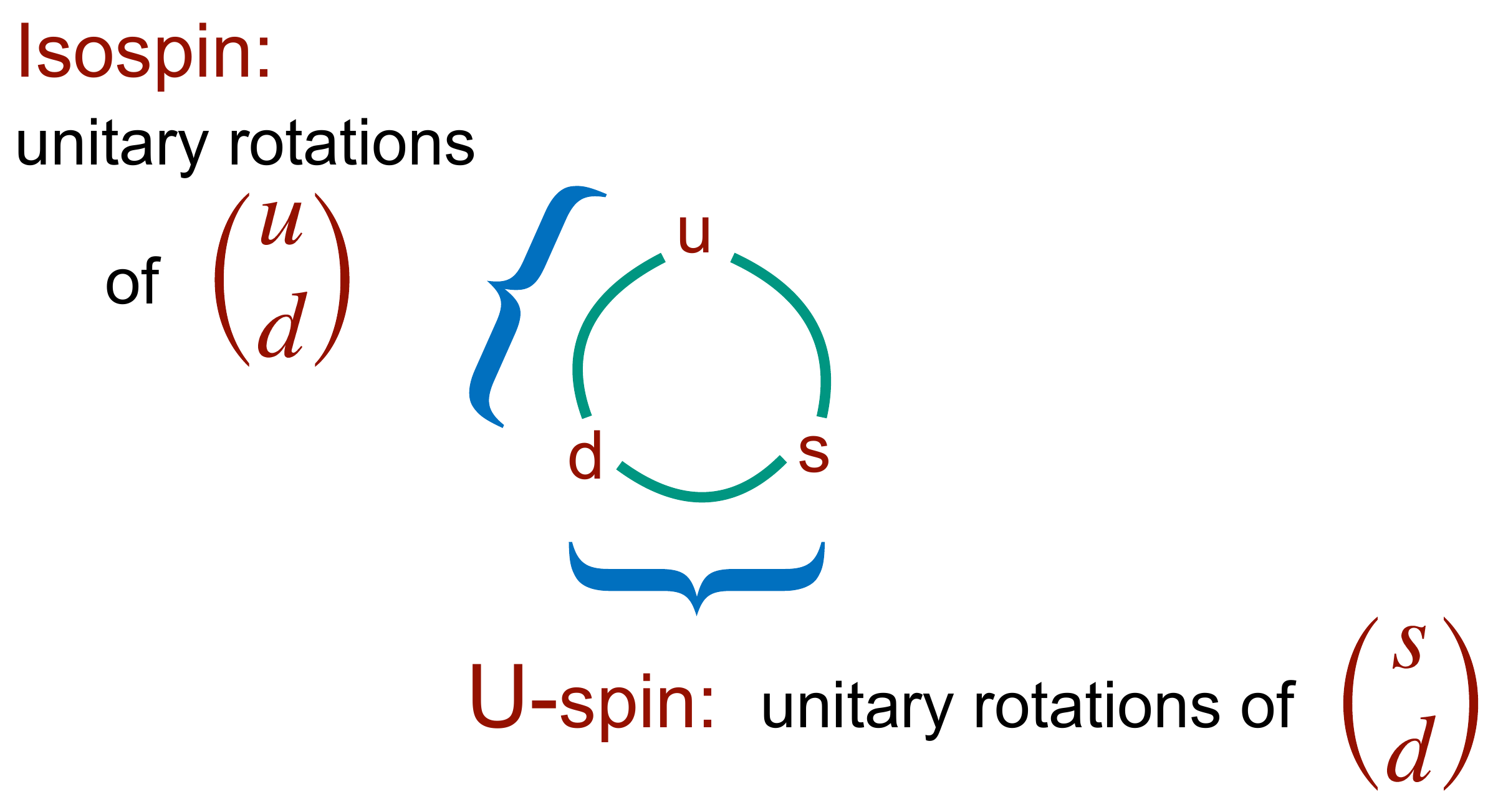} 
\caption{The $SU(3)_{\rm F}$ symmetry rotating the quark triplet
  $(u,d,s)^T$ would be an exact symmetry if the three quarks had the
  same mass. While isospin symmetry holds with an accuracy of
  about 2\%, U-spin symmetry is broken by 20-30\% due to
  $m_s\neq m_d$. 
  \label{fig:su3}}
~\\[-3mm]
\hrule
\end{figure}

There are no reliable methods to perform dynamical calculations for
exclusive hadronic decays of charmed hadrons. But it is possible to use
the approximate global symmetry $SU(3)_{\rm F}$ of QCD, which
corresponds to unitary rotations of the quark triplet $(u,d,s)^T$, to
relate the amplitudes of different decays to each other. The subscript
``F'', meaning ``flavour'', is added to distinguish $SU(3)_{\rm F}$ from
the QCD colour gauge symmetry $SU(3)_c$. The two most prominent $SU(2)$
subgroups of $SU(3)_{\rm F}$ correspond to the \emph{isospin}\ and
\emph{U-spin}\ subgroups explained in \fig{fig:su3}.  There is a long
history of global $SU(3)_{\rm F}$ analyses of hadronic two-body decays
of heavy hadrons. In Ref.~\cite{Gronau:1995hm} an analysis of branching
ratios of $B$ decays including linear (\ie\ first-order) $SU(3)_{\rm F}$
breaking has been presented. Corresponding analyses for $D$ decays can
be found in Refs.~\cite{Grossman:2012ry,Hiller:2012xm,
  Muller:2015lua,Bolognani:2024zno}.
In the practical implementation of $SU(3)_{\rm F}$ breaking
one treats the corresponding piece
$ H_{\cancel{\mathrm{SU(3)_F}} }= (m_s-m_d) \bar{s}{s}$ of the QCD
hamiltonian as a perturbation. (We do not consider isospin breaking
effects; in the $SU(3)_{\rm F}$ symmetry limit the three light quarks
have the common mass $m_d$.)
Including linear $SU(3)_{\rm F}$ breaking permits the reduction of the
intrinsic ${\cal O} (30\%) $ error of the predictions to an uncertainty
of ${\cal O} (10\%) $. Such global analyses involve theoretical building
blocks, \ie\ complex parameters entering the various decay amplitudes in
different combinations. Higher orders in the  $SU(3)_{\rm F}$ breaking
parameter $m_s-m_d$ bring more such parameters into the game, so that
successful predictions require good data on sufficiently many decay
modes. 

\section{$\mathbf{\eta}$-$\mathbf{\eta^\prime}$ mixing angle and
  $\mathbf{D\to P \eta^\prime}$ decays}
In this section I discuss the results of Ref.~\cite{Bolognani:2024zno}.

The pseudoscalar meson $SU(3)_{\rm F}$ octet comprises the states of
$\pi^\pm,\pi^0,K^\pm,K^0,\bar K{}^0$ and $\eta_8$. $SU(3)_{\rm F}$
breaking leads to mixing of the latter with the singlet $\eta_1$. This
feature is commonly parametrised in term of a mixing angle $\theta$ as
\begin{eqnarray}
\begin{aligned}
  |{\eta_8}\rangle
  &=\phantom{-}|\eta\rangle \cos\theta + |\eta^\prime\rangle\sin\theta \\
  |{\eta_1}\rangle
  &=-|\eta\rangle \sin\theta + |\eta^\prime\rangle\cos\theta
\end{aligned} \label{angle}
\end{eqnarray}
The mixing angle $\theta$ vanishes in the limit of exact $SU(3)_{\rm F}$
symmetry.  $SU(3)_{\rm F}$ breaking leads to non-zero off-diagonal terms
in the $\eta,\eta^\prime$ mass matrix and we define  $\theta$ as the
angle diagonalising this matrix.

\begin{figure}
\begin{center}%
\includegraphics[height=20mm]{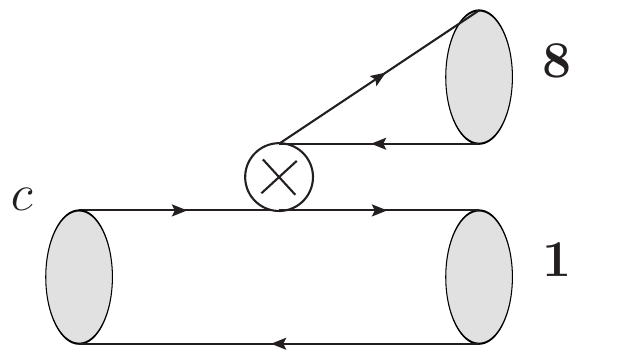}\hspace{10mm}
\includegraphics[height=20mm]{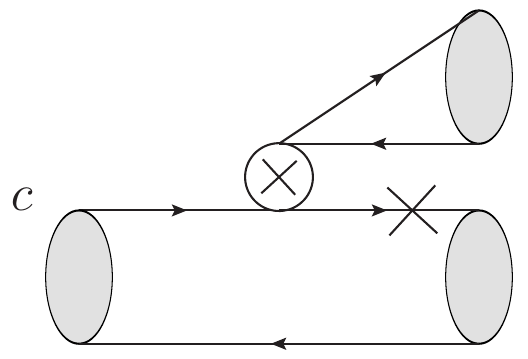}\hspace{5mm}
\includegraphics[height=20mm]{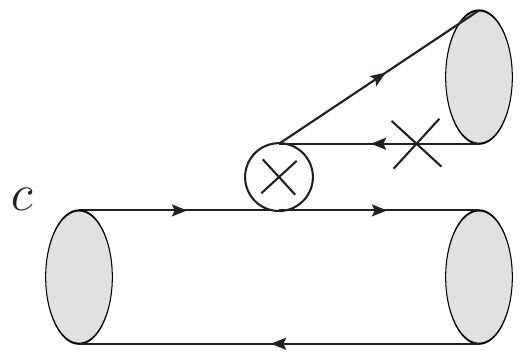}\hspace{5mm}
\includegraphics[height=20mm]{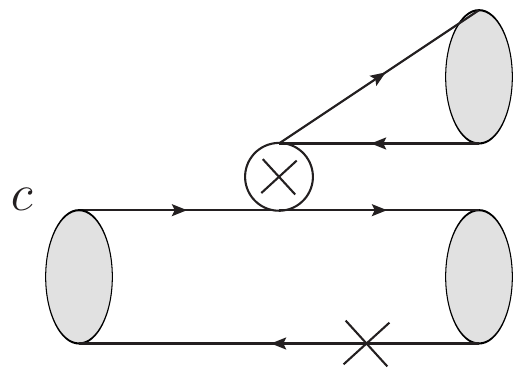}\hspace{5mm}
\end{center}

\caption{Sample topological amplitudes. The left diagram shows a
  $SU(3)_{\rm F}$ limit contribution for a $D$ meson decay into
  a singlet-octet final state. The three other diagrams depict
  first-order $SU(3)_{\rm F}$-breaking contributions, with the cross 
  on a strange-quark line indicating the Feynman rule for $
  H_{\cancel{\mathrm{SU(3)_F}} }$. 
    \label{fig:top}}
~\\[-3mm]
\hrule
\end{figure} 
Global $SU(3)_{\rm F}$ analyses of heavy hadron decays relate matrix elements
with $\pi$ or $K$ to those with $\eta_8$. To express  the latter in terms
of matrix elements with physical mesons, it is common practice to
use \eq{angle} schematically as
\begin{eqnarray}
\langle \eta \ldots | \ldots|\ldots\rangle &=& \langle \eta_8 \ldots |
  \ldots|\ldots\rangle\,\cos\theta \,-\, \langle \eta_1 \ldots |
  \ldots|\ldots\rangle\,\sin\theta,
\end{eqnarray}
with a similar expression for $\langle \eta^\prime \ldots |
\ldots|\ldots\rangle$ and to treat $\theta$ as a universal parameter.
However, such an approach is inconsistent. Specifying to our case of  
interest, we write the $D$ decay matrix elements as 
\begin{equation}
\begin{aligned}
  \langle P\eta | H | D \rangle 
  &=\, \cos\theta \langle P\eta_8  H | D \rangle  -
        \sin\theta \langle P \eta_1 |  H | D \rangle ,  
\nn 
 \langle P\eta^\prime | H | D \rangle 
  &=\, \sin\theta \langle P\eta_8  H | D \rangle^\prime     +
        \cos\theta \langle P \eta_1 |  H | D \rangle^\prime.     
\end{aligned}
        \label{mel}
\end{equation}
Yet these matrix elements are three-point functions and depend on 
kinematic variables built from the momenta $p_D$, $p_\eta$, and
$p_{\eta^\prime}$ of the three mesons involved. Since $\eta$ and
$\eta^\prime$ have different masses, $p_\eta^2\neq p_{\eta^\prime}^2$,
one concludes that 
\begin{eqnarray}
  \langle P\eta_8 | H | D \rangle ' &\neq& \langle P \eta_8 H |D\rangle \nn
  \langle P\eta_1 | H | D \rangle ' &\neq& \langle P \eta_1 H |D\rangle.  
\end{eqnarray}
An immediate consequence of this observation is that there is no point
in combining $D\to P \eta$ and $D\to P \eta^\prime$ decays into a common
analyses. However, one can still perform such a fit for  $D\to P
\eta^\prime$ decays alone or instead do this for all $D\to P \eta$
decays in conjunction with other $D$ decays into octet-octet final states.  

The mixing-angle problem was first addressed in the context of $\eta$
and $\eta^\prime$ decay constants
\cite{Leutwyler:1997yr,Feldmann:1998sh,Feldmann:1998vh} by introducing
different mixing angles for different decay modes. This approach cannot
be applied to global $SU(3)_{\rm F}$ analyses, in which the
$\ket{\eta_8}$ component of $\ket{\eta}$ must be related to $\ket{\pi}$
and $\ket{K}$. In our approach of Ref.~\cite{Bolognani:2024zno} we
instead insisted on a universal mixing angle $\theta$ defined solely
from the $\eta,\eta^\prime$ mass matrix. Thus $\theta$ in \eq{angle} is
solely defined in terms of the strong interaction, as opposed to
definitions employing electromagnetic or weak decay matrix
elements.\footnote{Also in lattice gauge theory the $\eta$-$\eta^\prime$
  mixing angle is defined via physical matrix elements and therefore
  process-dependent, see \eg\ \cite{Bali:2021qem}.}  The drawback of our
definition is that $\theta$ is not directly related to any physical
observable. In our analysis $\theta$ counts as first order in the
$SU(3)_{\rm F}$ breaking parameter and always appears together with
hadronic quantities parametrising first-order corrections to decay
matrix elements. Only the sum of the product of $\sin\theta$
with some $SU(3)_{\rm F}$-limit matrix elements and certain $SU(3)_{\rm F}$-breaking
corrections to these matrix elements is physical, so that $\theta$
cannot be determined from the global fit. 

We have performed a Frequentist statistical analysis using the branching
fractions of 
\begin{eqnarray*}
  && D^0\to\pi^0 \eta^\prime, \;\quad D^0\to\eta \eta^\prime,\quad\;\;
     D^+\to\pi^+ \eta^\prime,\quad  D_s^+\to K^+ \eta^\prime, \nn
  && D^0\to\bar K^0 \eta^\prime,\quad D_s^+\to\pi^+ \eta^\prime,
     \quad D^0\to K^0\eta^\prime, \quad D^+\to K^+\eta^\prime. \no
\end{eqnarray*}
The complex hadronic quantities serving as building blocks for 
the decay amplitudes are topological amplitudes
\cite{Wang:1980ac,Zeppenfeld:1980ex}, see \fig{fig:top}. The topological amplitudes
describing $SU(3)_{\rm F}$-breaking corrections from
$ H_{\cancel{\mathrm{SU(3)_F}} }= (m_s-m_d) \bar{s}{s}$ involve a cross
on a strange-quark line \cite{Gronau:1995hm,Muller:2015lua}. 
Alternatively, one can use the reduced matrix elements of the
Wigner-Eckart theorem as building blocks \cite{Grossman:2012ry,Hiller:2012xm}.  
In Ref.~\cite{Muller:2015lua} the mapping between these reduced matrix
elements and the topological amplitudes have been presented, showing the
equivalence of both approaches. 

\begin{figure}
\centering\includegraphics[width=0.9\textwidth]{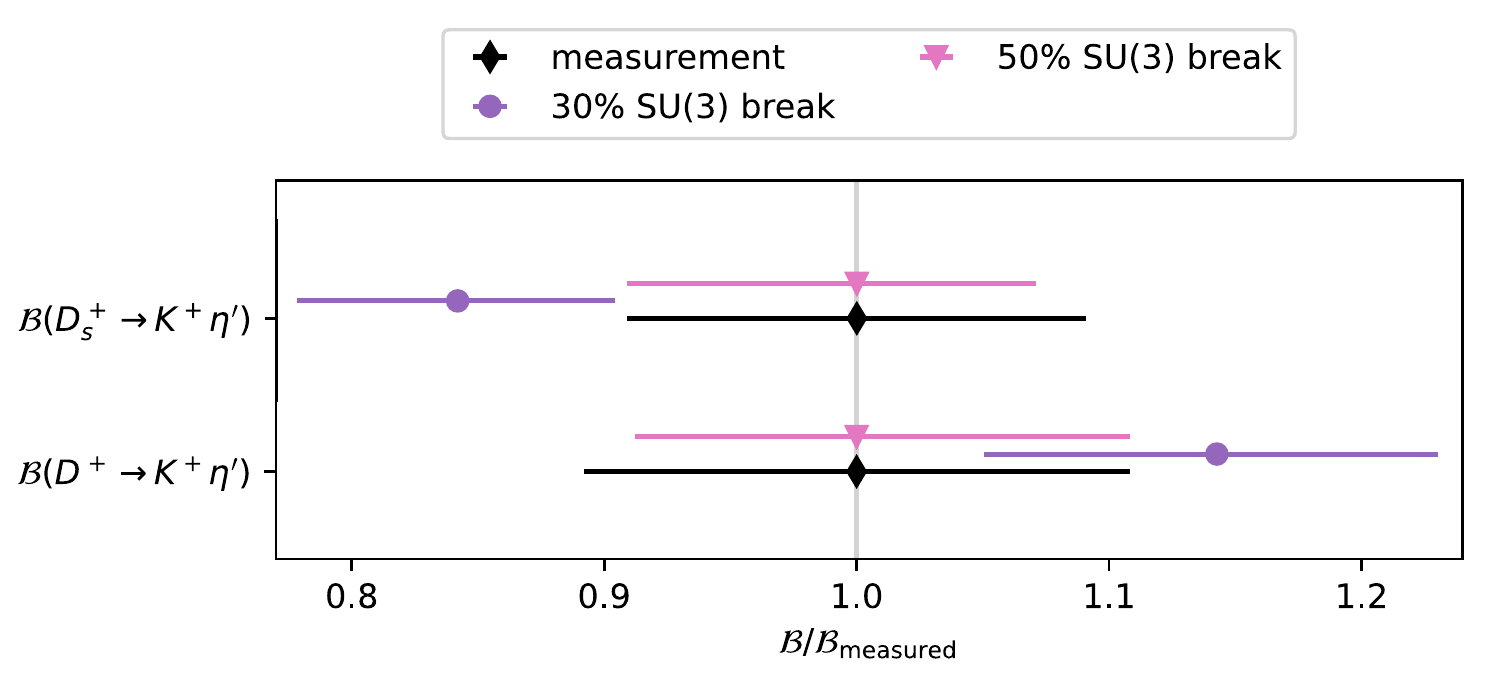}
\caption{If the size of the   $SU(3)_{\rm F}$-breaking contributions is
  limited to 30\%, the predictions for the two shown
  branching ratios deviate slightly from their measurements. If one
  permits 50\% $SU(3)_{\rm F}$ breaking, the global fit essentially
  reproduces the experimental input. See Ref.~\cite{Bolognani:2024zno}
  for details.  
  \label{fig:lh}}
~\\[-3mm]
\hrule
\end{figure}

The main results of our analyses are
\begin{itemize}\addtolength{\itemsep}{-5pt}
\item The global fit is consistent with  $\leq$30\%
  $SU(3)_{\rm F}$ breaking in the amplitudes.
\item The $SU(3)_{\rm F}$ limit is ruled out by 5.6$\sigma$. 
\item The fit predicts the branching fractions of $D_s^+\to
  K^+\eta^\prime$ and $D^+\to K^+\eta^\prime$
  by $\sim 1\sigma$ too low and too high, respectively, see \fig{fig:lh}. 
\end{itemize}

\section{CP asymmetries in hadronic two-body $\mathbf{D}$ decays}
In this section I discuss the results of Ref.~\cite{Iguro:2024uuw}.

For  $SU(3)_{\rm F}$ analyses it is useful to decompose the decay
amplitude $\mathcal{A}^{\mathrm{SCS}}$ of a $D\to PP^\prime$ or
$D\to PV$ decays in terms of U-spin representations as  \cite{Golden:1989qx}
\begin{eqnarray}
  \mathcal{A}^{\mathrm{SCS}} \equiv  \lambda_{sd} A_{sd}
\, - \, \frac{\lambda_b}{2} A_{b}
\end{eqnarray}
with 
\begin{eqnarray}
  \lambda_{sd} \;=\; \frac{\lambda_s-\lambda_d}{2} \simeq \lambda_s,
  &&\qquad\qquad 
     -\frac{\lambda_b}{2} \;=\;\frac{\lambda_s+\lambda_d}{2}.
     \label{asdb}
\end{eqnarray}
For $A_{sd}$ and $A_b$ are $|\Delta U|=1$ (triplet) and $\Delta U=0$
(singlet) transitions, respectively. Since $D^0$ carries no U-spin, in
$D^0$ decays these quantum numbers directly translate into those of the
final state.  \eq{asdb} translates to the more commonly used ``tree''
and ``penguin'' amplitudes as
\begin{eqnarray}
  \mbox{``tree''} \simeq A_{sd}, &&\qquad \qquad 
  \mbox{``penguin''} \simeq -\frac{A_b}2. \label{tpa}
\end{eqnarray}
In the SM direct CP violation stems from the interference of
$A_{sd}$ and $A_b$. The corresponding CP asymmetry for the decay
$D\to f$ reads
\begin{eqnarray}
  a_{CP}^{\mathrm{dir}}(D\to f) \;\equiv\;
  \frac{\Gamma(D\to f)-
       \Gamma(\bar D \to \bar f)}{\Gamma(D\to f)+ \Gamma(\bar D \to \bar f)}
  &\simeq&\mathrm{Im}\frac{\lambda_b}{\lambda_{sd}} \; 
  \mathrm{Im}\frac{A_b(D\to f)}{A_{sd}(D\to f)} .\label{cpdsm}
\end{eqnarray}
Here $\Gamma(D\to f)$ denotes the decay rate and $\bar f$  
is the CP-conjugate final state to $f$.
In \eqsto{asdb}{cpdsm} ``$\simeq$'' means that sub-leading terms 
in $\lambda_b/\lambda_{sd}$ have been neglected.

\subsection{$\mathbf{D^0\rightarrow\pi^+\pi^-}$ and
  $\mathbf{D^0\rightarrow K^+K^-}$}
\begin{figure}
  \includegraphics[width=0.9\textwidth]{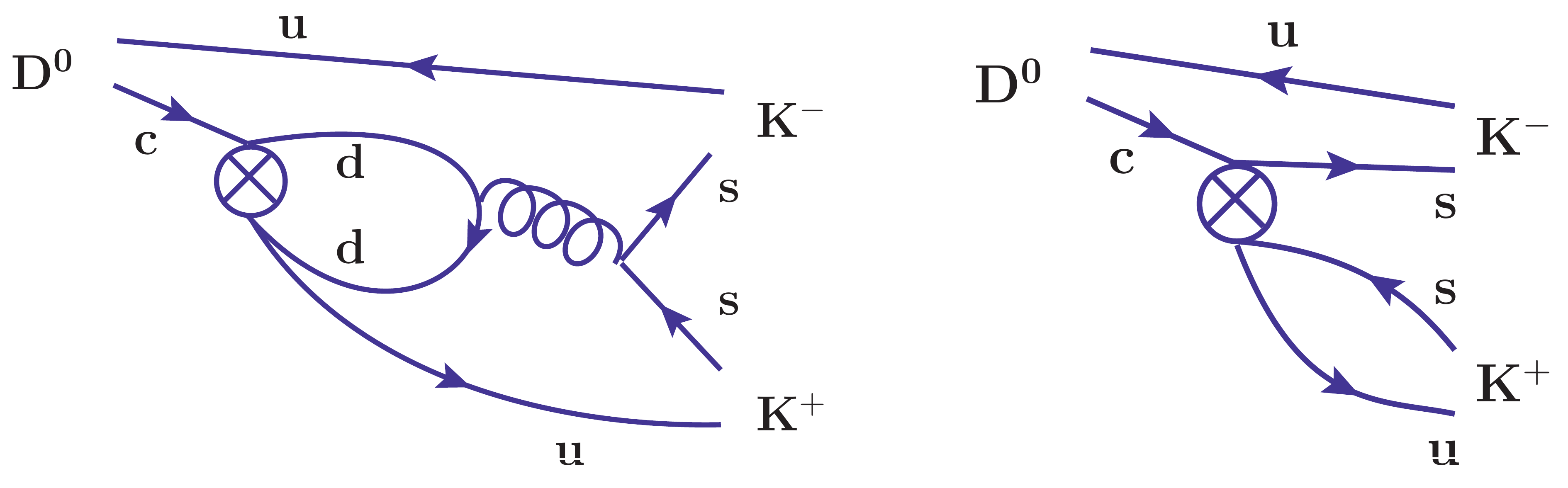}
  \caption{Dominant SM contributions to $A_b$ (left) and
    $A_{sd}$ (right). $a_{\rm CP} (D^0\to K^+K^-)$ is proportional to 
    $\mbox{Im}\, \frac{\lambda_d}{\lambda_s}=-\mbox{Im}\,\frac{\lambda_b}{\lambda_s}$
    and $\mbox{Im}\, \frac{A_b}{A_{sd}}$, see \eq{imbs}. A non-vanishing
    $\mbox{Im}\, \frac{A_b}{A_{sd}}$ requires rescattering, meaning that
    the intermediate $\bar u d\bar d u$ state is an on-shell $\pi\pi$
    or multi-hadron state which scatters into $K^+K^-$. 
    \label{fig:acpkk}}
~\\[-3mm]
\hrule
\end{figure}

On March 21, 2019, the LHCb collaboration announced the discovery of
charm CP violation through the measurement \cite{LHCb:2019hro}
\begin{eqnarray}
  \Delta a_{CP} \equiv \;\;a_{CP}^{\mathrm{dir}}(D^0\rightarrow K^+K^-)
  - a_{CP}^{\mathrm{dir}}(D^0\rightarrow\pi^+\pi^-)
   &=& (-15.4\pm 2.9)\cdot 10^{-4}. \label{acpexp}
\end{eqnarray}
LHCb has measured the time-integrated decays, so that $\Delta a_{CP}$
may contain a contamination from mixing-induced CP violation. This
potential contribution is much smaller than the error in \eq{acpexp} and
we omit the superscript ``dir'' occasionally.
In the U-spin limit $m_s=m_d$ one finds 
\begin{eqnarray}
  A_b(D^0\to K^+K^-)&=&A_b(D^0\to \pi^+\pi^-), \qquad
  A_{sd}(D^0\to K^+K^-)\;=\; - A_{sd}(D^0\to \pi^+\pi^-). \quad
\end{eqnarray}
so that
\begin{eqnarray}
\Delta a_{\rm CP} = 2a_{\rm CP} (D^0\to K^+K^-)=-2a_{\rm CP} (D^0\to \pi^+\pi^-)
\label{su3acp}
\end{eqnarray}
The interfering diagrams contributing to $a_{\rm CP} (D^0\to K^+K^-)$
are shown in \fig{fig:acpkk}.
The measurement exceeds the QCD light-cone sum rule prediction
\cite{Khodjamirian:2017zdu}
\begin{eqnarray}
  |\Delta a_{CP}| &\leq&   (2.0\pm 0.3)\cdot 10^{-4} \label{sr}
\end{eqnarray}
by a factor of 7.\footnote{After the LHCb measurement this calculation
  was critically reviewed and essentially confirmed, with a slightly
  weaker bound, $ |\Delta a_{CP}| \leq 3.6 \cdot 10^{-4}$
  \cite{Chala:2019fdb}.}  QCD sum rules constitute a sound method of
dynamical calculations of hadronic quantities, which has proven to yield
correct predictions for various observables in $B$ physics. In the field
of charm physics QCD sum rules are not well tested yet; a key feature of
this method is that the sum over certain hadronic contributions is
calculated in perturbation theory. Since $D$ mesons are lighter than
their beautiful counterparts, an individual resonance could dominate an
amplitude and this approach might fail in charm physics. In
Ref.~\cite{Schacht:2021jaz} it has been suggested that $U=0$ resonances
like $f_0(1710)$ and $f_0(1790)$ could enhance $A_b(D^0\to K^+K^-)$ and
$A_b(D^0\to \pi^+\pi^-)$. Thus the first puzzle of charm CP violation is
\begin{center}
  \parbox{0.7\textwidth}{\centering\emph{Is $ \Delta a_{CP} $ in \eq{acpexp} due
      to new physics or poorly understood QCD dynamics enhancing penguin
      amplitudes?}}
\end{center}

In 2022 LHCb has measured \cite{LHCb:2022lry}
\begin{eqnarray}
a_{CP}(D^0\to K^+K^-) &=& (7.7 \pm 5.7) \cdot 10^{-4} \label{kkexp}
\end{eqnarray}
entailing 
\begin{eqnarray}
a_{\rm CP} (D^0\to \pi^+\pi^- )&=& (23.1\pm 6.1) \cdot 10^{-4} \label{pipiexp}
\end{eqnarray}
from $ \Delta a_{CP} $ in \eq{acpexp}. From \eqsand{su3acp}{acpexp} we
conclude that one expects
$a_{CP}(D^0\to K^+K^-)= -a_{\rm CP} (D^0\to \pi^+\pi^-)= -(7.7\pm 1.5)
\cdot 10^{-4}$ in the limit of exact $SU(3)_{\rm F}$ symmetry. The
central values in \eqsand{kkexp}{pipiexp} are far away from these
$SU(3)_{\rm F}$ limit values, so that the situation is very different
from branching fractions for which $SU(3)_{\rm F}$ breaking is at the
nominal value of 30\% or smaller
\cite{Grossman:2012ry,Muller:2015lua,Bolognani:2024zno}.  The important
observations are
\begin{itemize}\addtolength{\itemsep}{-5pt}
\item \eq{kkexp} complies with the QCD sum rule calculation in
  \cite{Khodjamirian:2017zdu} at $1.1\sigma$. 
\item With future data $a_{CP}(D^0\to K^+K^-)$ in \eq{kkexp}
  must flip sign to comply with
  the approximate U-spin symmetry prediction
  $a_{\rm CP} (D^0\to K^+K^-) \approx -a_{\rm CP} (D^0\to \pi^+\pi^-)$. 
\end{itemize}
Thus the second puzzle of charm CP violation is
\begin{center}
  \parbox{0.7\textwidth}{\centering \emph{What causes the large
      violation of U-spin symmetry in $a_{\rm CP} (D^0\to K^+K^-)$
      vs. $a_{\rm CP} (D^0\to \pi^+\pi^-)$?}}
\end{center}
In view of the large error in \eq{kkexp} the second puzzle is not a
severe problem yet, a future shift of $a_{\rm CP} (D^0\to K^+K^-)$ by
2$\sigma$ (to a negative value) will eliminate the second puzzle
\cite{Schacht:2022kuj,Iguro:2024uuw}. If, however, we have to give up
U-spin symmetry, also explanations of $\Delta a_{\rm CP} $ in terms of
resonant enhancements \cite{Schacht:2021jaz} cannot be upheld, because
large U-spin breaking associated with these resonances would also unduly
enhance $|A_{sd}|$ and thereby branching fractions, spoiling the good
agreement of the latter with approximate $SU(3)_{\rm F}$ symmetry.

\begin{figure}
\centering\includegraphics[width=0.9\textwidth]{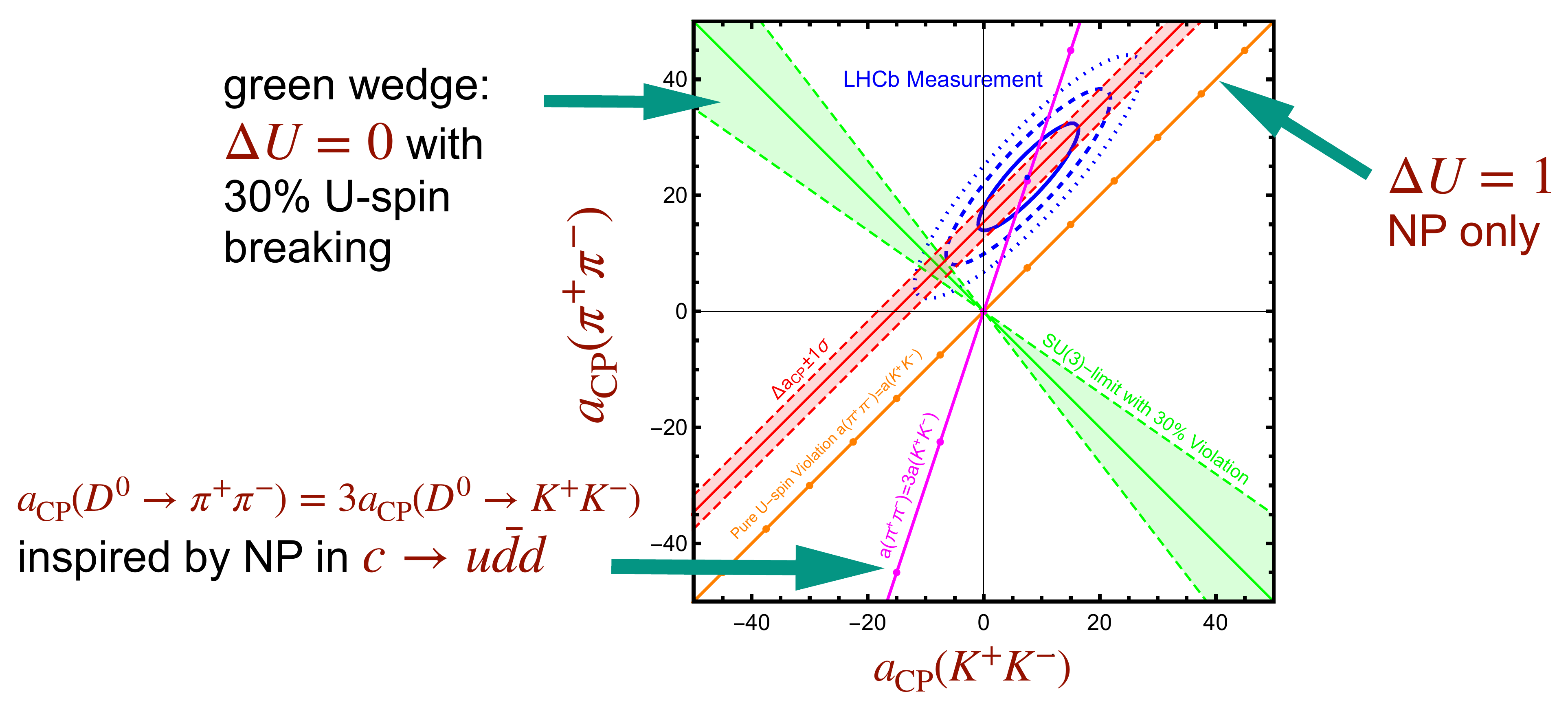}
\caption{The blue ellipses show the $1\sigma$, $2\sigma$, and
  $3\sigma$ ranges of \eqsand{acpexp}{kkexp}. The green wedge
  covers the SM (allowing an arbitrarily large  penguin amplitude $A_b$)
  and the case of $\Delta U=0$ new physics, permitting 30\% violation of
  U-spin symmetry. The orange line corresponds to $|\Delta U|=1$ new
  physics with no  $\Delta U=0$ contribution. The scenario
  $a_{\rm CP} (D^0\to \pi^+\pi^-)=3 a_{\rm CP} (D^0\to K^+ K^-)$
  explains the data. Plot from Ref.~\cite{Iguro:2024uuw}.
  \label{fig:plot}}
~\\[-3mm]
\hrule
\end{figure}

Postulating new physics in $|\Delta U|=1$ interactions solves the second
puzzle. A generic new $|\Delta U|=1$ interaction is proportional to
\begin{equation}
\ov{u}\Gamma c \left( \ov{d}\Gamma^\prime d - \ov{s}\Gamma^\prime s
\right), \label{np1}
\end{equation}
where the Dirac structures $\Gamma$ and $\Gamma^\prime$ need not be
specified for the presented symmetry-based analysis. The
generic $\Delta U=0$ interaction involves a ``+'' sign between the two
terms and may have a second term proportional to 
$\ov{u}\Gamma c\, \ov{u}\Gamma u$ with an independent coupling strength. 
Subsuming these cases into
\begin{eqnarray}
\mathcal{A}^{\mathrm{SCS}} &\equiv&  \lambda_{sd} A_{sd}
\, + \, a A_{\rm NP} \label{subs}
\end{eqnarray}
one finds, neglecting the SM penguin contribution,
\begin{eqnarray}
a_{CP}^{\mathrm{dir}} &=& -2\, \mathrm{Im}\frac{a}{\lambda_{sd}} \; 
  \mathrm{Im}\frac{A_{\rm NP}}{A_{sd}}
\end{eqnarray}
in analogy to \eq{cpdsm}. The two different cases come with different
relative signs in the $A_{\rm NP}$ amplitudes of U-spin related
decays. For example
$A_{\rm NP} (D^0\to K^+K^-) = -A_{\rm NP} (D^0\to \pi^+ \pi^-)$ in the
U-spin limit for $|\Delta U|=1$ new physics, while these amplitude are
the same for the $\Delta U=0$ case. 
The essential features of these scenarios is that   $\Delta U=0$ new
physics is indistinguishable from an enhanced SM penguin amplitude
$A_b$, while   $|\Delta U|=1$ new physics leads to different relations
between CP asymmetries, in particular
\begin{eqnarray}
  a_{\rm CP} (D^0\to K^+K^-)&=& a_{\rm CP} (D^0\to \pi^+\pi^-).
                                \label{cpu1}
\end{eqnarray}
in the U-spin limit if  $|\Delta U|=1$ new physics dominates over the SM
$A_b$ amplitude.  \fig{fig:plot} confronts the measurements in
\eqsand{acpexp}{kkexp} with the SM and new-physics scenarios.
Clearly, \eq{cpu1} forbids an explanation in terms of $|\Delta U|=1$ new
physics alone and the corresponding orange line in \fig{fig:plot} does
not intersect the experimentally allowed range. We may speculate about
new physics in $c\to u \bar d d$ decays, which contributes to
$D^0\to \pi^+ \pi^-$ at tree level and to $D^0\to K^+ K^-$ through a
penguin loop. The latter is suppressed by a colour factor of $1/N_c=1/3$
w.r.t. the tree amplitude. If furthermore the strong phases of tree and
penguin amplitude are similar, one will find
$a_{\rm CP} (D^0\to \pi^+\pi^-) \approx 3 a_{\rm CP} (D^0\to K^+ K^-)$
which fits the data well, since the purple line in \fig{fig:plot} spikes the center of the error
ellipses. The $c\to u \bar d d$ case is an example of a scenario with
both $|\Delta U|=1$ and $\Delta U=0$ new physics.  Several authors have
considered specific new-physics models to address \eqsand{acpexp}{kkexp}
\cite{Chala:2019fdb,Dery:2019ysp,Calibbi:2019bay,Bause:2020obd,
  Buras:2021rdg,Bause:2022jes}.

To summarise:
\begin{itemize}\addtolength{\itemsep}{-5pt}
\item If $a_{\rm CP} (D^0\to \pi^+\pi^-)$ is governed by the SM\ldots
  \begin{itemize}\addtolength{\itemsep}{-5pt}
 \item[\ldots\hspace{-1ex}]the QCD sum rule calculation does not work \emph{and}
 \item[\ldots\hspace{-1ex}]either U-spin symmetry fails for
   $A_b$ or in future measurements $a_{\rm CP} (D^0\to K^+ K^-)$
   will move down by 2$\sigma$ from the value in \eq{kkexp} and flip sign.
\end{itemize}
\item  If $a_{\rm CP} (D^0\to \pi^+\pi^-)$ is dominated by new physics\ldots
\begin{itemize}\addtolength{\itemsep}{-5pt}
\item[\ldots\hspace{-1ex}]the new-physics contribution necessarily has a $|\Delta
  U|=1$ piece \emph{and} 
\item[\ldots\hspace{-1ex}]there is  an additional  $\Delta U|=0$ new-physics
   contribution or some enhancement of $A_b$ over the sum rule prediction. 
\end{itemize}
\end{itemize}

\subsection{Sum rules for CP asymmetries}
In order to shed light on the two puzzles mentioned above one
must measure CP asymmetries in as many decays of charmed hadrons as
possible. 
Employing U-spin symmetry we have derived sum rules between direct CP
asymmetries of SCS $D$ meson decays in Ref.~\cite{Iguro:2024uuw}
for the generic $|\Delta   U|=1$ and $\Delta U=0$ new-physics
scenarios. Such sum rules have been derived  for the SM (and thereby
also for $\Delta U=0$ new physics)
in Refs.~\cite{Grossman:2012ry,Grossman:2013lya,Muller:2015rna}. 
Our findings for the $\Delta U=0$ sum rules comply with those of
Ref.~\cite{Grossman:2013lya}, but we have found one extra sum rule each
for $D\to PP^\prime$  and $D\to PV$  decays. The sum rules for
the $|\Delta U|=1$ new-physics scenario are derived for the case of \eq{np1}.
The sum rules involve between two and ten CP asymmetries. 
Here I present only those which are easiest to test at LHCb.
To this end it is useful to define 
\begin{eqnarray}
  A_{CP} (D\to f) &\equiv& 2 \bar\Gamma(D\to f)  a_{CP} (D\to f)
                           \;=\; \Gamma(D\to f) -\Gamma(\bar D\to \bar f) 
\end{eqnarray}
with $\bar\Gamma(D\to f)$  being the average of $\Gamma(D\to f)$ and
$\Gamma(\bar D\to \bar f)$. 

We find
\begin{eqnarray*}
     \begin{array}{|c|c|}\hline
       \Delta U=0
       &       |\Delta U|=1 \\\hline
      A_{\rm CP}(D^0\to K^+K^-)+A_{\rm CP}(D^0\to
       \pi^+\pi^-)=0
        &
          A_{\rm CP}(D^0\to K^+K^-)-A_{\rm CP}(D^0\to
          \pi^+\pi^-)=0 \\
       A_{\rm CP}(D_s^+\to K^0\pi^+)+A_{\rm CP}(D^+\to\bar
       K{}^0 K^+)=0&
                     A_{\rm CP}(D_s^+\to K^0\pi^+)-A_{\rm CP}(D^+\to\bar K{}^0 K^+)=0
       \\\hline
       A_{\rm CP}(D^0\to K^0 \bar K^{*0})+A_{\rm CP}(D^0\to
       \bar K^0 K^{*0})=0&
                        A_{\rm CP}(D^0\to K^0 \bar
                           K^{*0})-A_{\rm CP}(D^0\to \bar K^0
                           K^{*0})=0 \\
       A_{\rm CP}(D_s^+\to K^{*0}\pi^+) + A_{\rm CP}(D^+\to
       \bar K^{*0}K^+)=0&
                          A_{\rm CP}(D_s^+\to K^{*0}\pi^+)
                          - A_{\rm CP}(D^+\to \bar K^{*0}K^+)=0 \\\hline
     \end{array}
\end{eqnarray*}
For the remaining sum rules see Ref.~\cite{Iguro:2024uuw}.  To explain
\eqsand{acpexp}{kkexp} we need both $\Delta U=0$ and $|\Delta U|=1$ 
contributions.  Since $A_{CP}$ is linear in
$A_{\rm NP}= A_{\rm NP}^{\Delta U=0} +A_{\rm NP}^{|\Delta U|=1}$ (and
$A_b$), we can write
\begin{eqnarray}
  A_{CP} (D\to f) &=& A_{CP} (D\to f)^{\Delta U=0} +A_{CP} (D\to f)^{|\Delta U|=1}
\end{eqnarray}
with the first and second term obeying the sum rule of the first and
second column of the table, respectively. Solving these equations for
the decays in the first row, one finds 
\begin{eqnarray}
A_{CP} (D\to \pi^+\pi^-)^{\Delta U=0} & =&-
  A_{CP} (D\to K^+K^-)^{\Delta U=0} \nn
  &= &\phantom{-}
    \frac{A_{CP} (D\to \pi^+\pi^-)-A_{CP} (D\to K^+K^-)}2 \nn
 A_{CP} (D\to \pi^+\pi^-)^{|\Delta U|=1} & =&  
  A_{CP} (D\to K^+K^-)^{|\Delta U|=1} \nn
  &= &
    \frac{A_{CP} (D\to \pi^+\pi^-)+A_{CP} (D\to K^+K^-)}2                                        
\end{eqnarray}
Taking $A_{CP} (D\to \pi^+\pi^-)=2 A_{CP} (D\to K^+K^-)$ as a numerical
example one finds
$ A_{CP} (D\to \pi^+\pi^-)^{|\Delta U|=1}=3 A_{CP} (D\to K^+K^-)^{\Delta
  U=0}$, which thus fixes the relative size of the contributions from
$A_{CP}^{|\Delta U|=1}$ and $A_b+ A_{\rm CP}^{\Delta U=0}$ in the two
studied decay modes. Since $SU(3)_{\rm F}$ symmetry is approximate one
should include an uncertainty of order 30\%.

An experimental advantage of $ \Delta a_{CP}$ compared to the individual
CP asymmetries is the cancellation of the $D^0$ vs.\ $\bar D^0$
production asymmetry from the measured quantity. In
Ref.~\cite{Iguro:2024uuw} we have  proposed similar combinations for
the CP asymmetries entering our sum rules. For example, instead of
measuring $a_{\rm CP}(D^+\to \bar K^{*0}K ^+)$ one could measure
\begin{eqnarray}
 \Delta  a_{\rm CP,9}(D^+)
  &\equiv & a_{\rm CP}(D^+\to \bar K^{*0}K ^+) - a_{\rm CP}(D^+\to \bar
            K^{*0} \pi^+) \label{dacp}
\end{eqnarray}
The decay $D^+\to \bar K^{*0} \pi^+$ has no SM CP violation and also
new-physics contributions can barely compete with the CF tree amplitude
governing this decay. So it is safe to assume that the subtracted CP
asymmetry in \eq{dacp} is much smaller than
$a_{\rm CP}(D^+\to \bar K^{*0}K ^+)$.  Playing the same game for
$a_{\rm CP}(D_s^+\to K^{*0}\pi^+)$ requires the subtraction of
$a_{\rm CP}(D_s^+\to K^0K^{*+})$ which, however, is DCS. DCS decays have
no penguin amplitude but there could be a tree contribution from new
physics \cite{Bergmann:1999pm}. A future measurement of a non-zero
$ \Delta a_{\rm CP,5}(D_s^+)\equiv a_{\rm CP}(D_s^+\to K^{*0}\pi^+) -
a_{\rm CP}(D_s^+\to K^0K^{*+})$ will trigger an interesting discussion
on which of the two decays is responsible for the finding. As a final
remark, the production asymmetry is not an issue in the $\Delta U=0$  
scenario for $a_{\rm CP}(D^0\to K^0 \bar K^{*0})$ and $a_{\rm CP}(D^0\to \bar K^0
K^{*0})$, since these CP asymmetries can be measured without flavour
tagging \cite{Nierste:2017cua}. But searching  for $|\Delta U|=1$ new physics
does require to distinguish $D^0$ from $\bar D^0$ decays, because
$A_{\rm NP}^{|\Delta U|=1}$ drops out from
$a_{\rm CP}\left(\DorDbar\to K^0 \bar K^{*0}\right)$, where
$\DorDbar$ denotes an untagged $D$ or $\bar D$ meson. 

\section{Summary}
\begin{enumerate}\addtolength{\itemsep}{-5pt}
\item A universal $\eta$-$\eta^\prime$ mixing angle defined through
  unitary rotations of matrix elements with $\eta_8$ and $\eta_1$ is
  known since 27 years to be ill-defined. It is nevertheless commonly
  used in global $SU(3)_{\rm F}$ analyses of $D$ or $B$ decay data.
\item We have devised a consistent treatment of $\eta\!-\!\eta^\prime$
  mixing, which permits a global analysis of $D\to P \eta^\prime$
  or $D\to P \eta$ data, while it is not possible to relate the former
  decays to the latter. 
\item A global fit to the branching ratios of
  $D^0\to\pi^0 \eta^\prime,D^0\to\eta \eta^\prime, D^+\to\pi^+
  \eta^\prime, D_s^+\to K^+ \eta^\prime, D^0\to\bar K^0 \eta^\prime,
  D_s^+\to\pi^+ \eta^\prime, D^0\to K^0\eta^\prime$, and $D^+\to
  K^+\eta^\prime$ complies with $\leq$30\%
  $SU(3)_{\rm F}$ breaking, with slight tensions in
  $ D_s^+\to K^+\eta^\prime$ and $D^+\to K^+\eta^\prime$.
\item The LHCb measurements $\Delta a_{CP} = (-15.4\pm 2.9)\cdot
  10^{-4}$
  and $a_{CP}(D^0\to K^+K^-) = (7.7 \pm 5.7) \cdot 10^{-4}$ are not
  consistent with the SM if U-spin symmetry holds approximately.
\item
  New physics explanations involve a $|\Delta U|=1$
  amplitude (with a different phase than $\arg(\pm V_{cs}^* V_{us})$) and a
  $\Delta U=0$ amplitude (SM or NP) as well. 
\item One can check this in the future in other decay modes in which
  CP asymmetries are not yet measured to be non-zero. To this end we
  have proposed sum rules between CP asymmetries. 
\item Especially interesting for LHCb are sum rules relating
  $a_{\rm CP}(D_s^+\to K^0\pi^+)$  to
  $a_{\rm CP}(D^+\to \bar K^0K^+)$ as well as sum rules involving CP
  asymmetries in $D^0\to K^0 \bar K^{*0}$, $D^0\to \bar K^0 K^{*0}$,
  $D_s^+\to K^{*0}\pi^+$, and $D^+\to \bar K^{*0}K^+$.
\item  In an experimental analysis one may choose to study differences
   like  $\Delta  a_{\rm CP,9}(D^+)
  \equiv  a_{\rm CP}(D^+\to \bar K^{*0}K ^+) - a_{\rm CP}(D^+\to \bar
            K^{*0} \pi^+)$ to eliminate production asymmetries. 
\end{enumerate}
Finally I mention the parallel talks on charm physics at this
conference:\\
Eleftheria Solomonidi, \emph{Implications of cascade topologies for rare
  charm decays and CP violation}, which is a theory talk,\\
Luca Balzani, \emph{Particle-antiparticle asymmetries in hadronic charm
  decays at LHCb} covering $D -\bar D$ mixing and CP violation, and\\
Marco Colonna, \emph{Rare charm decays at LHCb}, discussing $D\to h
h^\prime e^+e^-$ and more. 

\section*{Acknowledgements}
I thank the organisers of \emph{DISCRETE 2024}\ for their invitation to
give this plenary talk. I am grateful for the enjoyable collaborations
with Carolina Bolognani, Syuhei Iguro, Emil Overduin, Stefan Schacht,
Maurice Sch\"u\ss ler, and Keri Vos on the presented work in
Refs.~\cite{Bolognani:2024zno,Iguro:2024uuw}.

\end{document}

%% file: def_colours.tex
\definecolor{GreenYellow}   {cmyk}{0.15,0,0.69,0}
\definecolor{Yellow}        {cmyk}{0,0,1,0}
\definecolor{Goldenrod}     {cmyk}{0,0.10,0.84,0}
\definecolor{Dandelion}     {cmyk}{0,0.29,0.84,0}
\definecolor{Apricot}       {cmyk}{0,0.32,0.52,0}
\definecolor{Peach}         {cmyk}{0,0.50,0.70,0}
\definecolor{Melon}         {cmyk}{0,0.46,0.50,0}
\definecolor{YellowOrange}  {cmyk}{0,0.42,1,0}
\definecolor{Orange}        {cmyk}{0,0.61,0.87,0}
\definecolor{BurntOrange}   {cmyk}{0,0.51,1,0}
\definecolor{Bittersweet}   {cmyk}{0,0.75,1,0.24}
\definecolor{RedOrange}     {cmyk}{0,0.77,0.87,0}
\definecolor{Mahogany}      {cmyk}{0,0.85,0.87,0.35}
\definecolor{Maroon}        {cmyk}{0,0.87,0.68,0.32}
\definecolor{BrickRed}      {cmyk}{0,0.89,0.94,0.28}
\definecolor{Red}           {cmyk}{0,1,1,0}
\definecolor{OrangeRed}     {cmyk}{0,1,0.50,0}
\definecolor{RubineRed}     {cmyk}{0,1,0.13,0}
\definecolor{WildStrawberry}{cmyk}{0,0.96,0.39,0}
\definecolor{Salmon}        {cmyk}{0,0.53,0.38,0}
\definecolor{CarnationPink} {cmyk}{0,0.63,0,0}
\definecolor{Magenta}       {cmyk}{0,1,0,0}
\definecolor{VioletRed}     {cmyk}{0,0.81,0,0}
\definecolor{Rhodamine}     {cmyk}{0,0.82,0,0}
\definecolor{Mulberry}      {cmyk}{0.34,0.90,0,0.02}
\definecolor{RedViolet}     {cmyk}{0.07,0.90,0,0.34}
\definecolor{Fuchsia}       {cmyk}{0.47,0.91,0,0.08}
\definecolor{Lavender}      {cmyk}{0,0.48,0,0}
\definecolor{Thistle}       {cmyk}{0.12,0.59,0,0}
\definecolor{Orchid}        {cmyk}{0.32,0.64,0,0}
\definecolor{DarkOrchid}    {cmyk}{0.40,0.80,0.20,0}
\definecolor{Purple}        {cmyk}{0.45,0.86,0,0}
\definecolor{Plum}          {cmyk}{0.50,1,0,0}
\definecolor{Violet}        {cmyk}{0.79,0.88,0,0}
\definecolor{RoyalPurple}   {cmyk}{0.75,0.90,0,0}
\definecolor{BlueViolet}    {cmyk}{0.86,0.91,0,0.04}
\definecolor{Periwinkle}    {cmyk}{0.57,0.55,0,0}
\definecolor{CadetBlue}     {cmyk}{0.62,0.57,0.23,0}
\definecolor{CornflowerBlue}{cmyk}{0.65,0.13,0,0}
\definecolor{MidnightBlue}  {cmyk}{0.98,0.13,0,0.43}
\definecolor{NavyBlue}      {cmyk}{0.94,0.54,0,0}
\definecolor{RoyalBlue}     {cmyk}{1,0.50,0,0}
\definecolor{Blue}          {cmyk}{1,1,0,0}
\definecolor{Cerulean}      {cmyk}{0.94,0.11,0,0}
\definecolor{Cyan}          {cmyk}{1,0,0,0}
\definecolor{ProcessBlue}   {cmyk}{0.96,0,0,0}
\definecolor{SkyBlue}       {cmyk}{0.62,0,0.12,0}
\definecolor{Turquoise}     {cmyk}{0.85,0,0.20,0}
\definecolor{TealBlue}      {cmyk}{0.86,0,0.34,0.02}
\definecolor{Aquamarine}    {cmyk}{0.82,0,0.30,0}
\definecolor{BlueGreen}     {cmyk}{0.85,0,0.33,0}
\definecolor{Emerald}       {cmyk}{1,0,0.50,0}
\definecolor{JungleGreen}   {cmyk}{0.99,0,0.52,0}
\definecolor{SeaGreen}      {cmyk}{0.69,0,0.50,0}
\definecolor{Green}         {cmyk}{1,0,1,0}
\definecolor{ForestGreen}   {cmyk}{0.91,0,0.88,0.12}
\definecolor{PineGreen}     {cmyk}{0.92,0,0.59,0.25}
\definecolor{LimeGreen}     {cmyk}{0.50,0,1,0}
\definecolor{YellowGreen}   {cmyk}{0.44,0,0.74,0}
\definecolor{SpringGreen}   {cmyk}{0.26,0,0.76,0}
\definecolor{OliveGreen}    {cmyk}{0.64,0,0.95,0.40}
\definecolor{RawSienna}     {cmyk}{0,0.72,1,0.45}
\definecolor{Sepia}         {cmyk}{0,0.83,1,0.70}
\definecolor{Brown}         {cmyk}{0,0.81,1,0.60}
\definecolor{Tan}           {cmyk}{0.14,0.42,0.56,0}
\definecolor{Gray}          {cmyk}{0,0,0,0.50}
\definecolor{Black}         {cmyk}{0,0,0,1}
\definecolor{White}         {cmyk}{0,0,0,0}